\documentclass{moriond}

\usepackage{slashed}
\usepackage{amsmath}
\usepackage{amssymb}
\usepackage{xspace}

\def\be{\begin{equation}}
\def\ee{\end{equation}}
\def\bea{\begin{eqnarray}}
\def\eea{\end{eqnarray}}

\newcommand{\Photo}{}
\newcommand{\gambit}{\textsf{GAMBIT}\xspace}

\newcommand{\Q}[2]{
  \if\relax\detokenize{#2}\relax
    \mathcal{Q}_{#1}
  \else
    \mathcal{Q}_{#1}^{(#2)}
  \fi
}

\newcommand{\C}[2]{
  \if\relax\detokenize{#2}\relax
    \mathcal{C}_{#1}
  \else
    \mathcal{C}_{#1}^{(#2)}
  \fi
}

\newcommand{\La}{{\rm \Lambda}}

\begin{document}
\vspace*{4cm}
\title{BSM global fits with GAMBIT:\\ a Dark Matter EFT fit}

\author{Tom\'as E. Gonzalo, on behalf of the GAMBIT Community}

\address{Institute for Theoretical Particle Physics and Cosmology (TTK),\\ RWTH Aachen University, D-52056 Aachen, Germany}

\maketitle
\abstracts{In this conference paper I present the first full global fit of a dark matter effective field theory with the global fitting framework \gambit. I show the results of exhaustive parameter space explorations of the effective dark matter model, including a general set of operators up to dimension 7, and using the most up-to-date constraints from direct and indirect detection of dark matter, relic abundance requirements and collider searches for dark matter candidates.}

\section{Introduction}

The nature of dark matter (DM) and its interactions has been an extensively explored subject for a long time. Due to the large landscape of plausible models of DM, it is convenient to parametrise the DM interactions with Standard Model (SM) particles in a model-agnostic way, using the Effective Field Theory (EFT) approach~\cite{DMEFT}. EFT models often have vast parameter spaces and thus smart sampling strategies are needed to fully explore them~\cite{AbdusSalam:2020rdj}. One of the most powerful tools for such study is the \gambit framework~\cite{Athron:2017ard}. \gambit, the Global And Modular BSM Inference Tool, is an open-source global fitting software able to perform statistical fits on a variety of BSM models~\cite{Athron:2018vxy,Athron:2018hpc,Chrzaszcz:2019inj,Athron:2020maw,Stocker:2020nsx}, and thus it is perfectly suited for this purpose. 

\section{Dark Matter Effective Field Theory}

The most generic interaction lagrangian for an effective field theory of DM can be written as $\mathcal{L}_{\rm{int}} = \sum_{a,d} (\C{a}{d}/\La^{d-4}) \Q{a}{d}$, where $\Q{a}{d}$ is the interaction operator, $d$ its dimension, $\C{a}{d}$ the dimensionless Wilson coefficient (WC), and $\La$ is the scale of new physics. In this study we focus exclusively on the interactions between DM particles ($\chi$) and quarks and gluons, as they are the most relevant interactions for searches of DM. Hence, the effective operators we consider are
\begin{equation} 
 \scriptsize{
 \label{dim6efts}
 \notag \begin{array}{llllll}
  \Q{1,q}{6} &= (\overline\chi \gamma_\mu \chi)(\overline{q} \gamma^\mu q)\,, & \Q{1}{7} &= \frac{\alpha_s}{12\pi}(\overline\chi \chi)G^{a\mu\nu}G^a_{\mu\nu}\,, & \Q{5,q}{7} &= m_q(\overline\chi \chi)( \overline{q} q )\,,\\
  \Q{2,q}{6} &= (\overline\chi \gamma_\mu \gamma_5 \chi)(\overline{q} \gamma^\mu q)\,, & \Q{2}{7} &= \frac{\alpha_s}{12\pi}(\overline\chi i\gamma_5 \chi)G^{a\mu\nu}G^a_{\mu\nu}\,, & \Q{6,q}{7} &= m_q(\overline\chi i\gamma_5 \chi)( \overline{q} q )\,,\\
  \Q{3,q}{6} &= (\overline\chi \gamma_\mu \chi)(\overline{q} \gamma^\mu \gamma_5 q)\,, & \Q{3}{7} &= \frac{\alpha_s}{8\pi}(\overline\chi \chi)G^{a\mu\nu}\widetilde{G}^a_{\mu\nu}\,, & \Q{7,q}{7} &= m_q(\overline\chi \chi)( \overline{q} i\gamma_5 q )\,,  \\
  \Q{4,q}{6} &= (\overline\chi \gamma_\mu \gamma_5 \chi)(\overline{q} \gamma^\mu \gamma_5 q)\,, & \Q{4}{7} &= \frac{\alpha_s}{8\pi}(\overline\chi i\gamma_5 \chi)G^{a\mu\nu}\widetilde{G}^a_{\mu\nu}\,, & \Q{8,q}{7} &= m_q(\overline\chi i\gamma_5 \chi)( \overline{q} i\gamma_5 q )\,, \\
  &&&&\Q{9,q}{7} &= m_q(\overline\chi \sigma^{\mu\nu} \chi)( \overline{q} \sigma_{\mu\nu} q )\,, \\
  &&&&\Q{10,q}{7} &= m_q(\overline\chi i\sigma^{\mu\nu}\gamma_5 \chi)( \overline{q} \sigma_{\mu\nu} q )\,. \\
 \end{array}
 }
\end{equation}

Many of the constraints, such as direct detection, are computed using non-relativistic operators, at lower scales. The running of the relativistic operators above from the input scale ($\La$), with mixing and threshold effects included, as well as the matching to the non-relativistic operators, is performed per parameter point by \textsf{DirectDM} v2.2.0~\cite{Bishara:2017nnn}. Throughout this study we consider $\C{a}{d}$ and $\La$ as independent parameters. This allows the study to use weaker bounds but with larger range of validity in $\La$, as opposed to naive approaches that only vary $C/\Lambda^{d-4}$. Therefore, the free parameters in this model are the 6- and 7-dimensional WCs, $\C{a}{6}$ and $\C{a}{7}$, the new physics scale $\La$, and the DM mass $m_\chi$. In addition, a set of nuisance parameters are varied simultaneously corresponding to the DM halo profile, SM masses and nuclear parameters, amounting to a total of 24 scan parameters.

\section{Likelihoods and Constraints}

The effective operators described above predict potentially strong interactions between DM and the particles in the SM. Hence, the parameters of the EFT model, namely the WCs, the DM mass, $m_\chi$ and new physics scale $\La$, are strongly constrained by the various searches for DM. The specific set of constraints employed in this study is as follows:
\\

\noindent \textit{Direct detection}

\noindent DM particles from the Galactic halo can scatter off nuclei. Various direct detection experiments with ultra-pure targets can detect such rare processes, and thus enforce strong constraints on the interaction cross-section of DM particles. The direct detection experiments considered in this study are CDMSlite, CRESST-II and -III, DarkSide 50, LUX 2016, PICO-60, PandaX 2016 and 2017, and XENON1T. We use \textsf{DirectDM} to compute the value of the non-relativistic WCs, including all running and mixing effects, and \textsf{DDCalc} v2.2.0~\cite{Workgroup:2017lvb} to calculate the likelihood for each of the relevant direct detection experiments.
\\

\noindent \textit{Relic abundance}

\noindent The precise measurement of the relic abundance of DM by \textit{Planck}, $\Omega_{\rm DM}h^2 = 0.120 \pm 0.001$\cite{}, severely constrains the annihilation cross-section of DM particles in the Early Universe. As a conservative approach, we assume that the DM particles studied here do not constitute all of the observed DM, and thus we allow parameter combinations were DM is underabundant, with a DM fraction $f_\chi$. We use this DM fraction to scale the direct and indirect detection signals. We compute the relic density (RD) using \textsf{DarkSUSY} v6.2.2~\cite{Bringmann:2018lay}, from tree-level cross-sections calculated with \textsf{CalcHEP} v3.6.27~\cite{Belyaev:2012qa}, from the interactions generated with \textsf{GUM}~\cite{Gonzalo:2021cnq}. In order to preserve EFT validity, we ignore parameter points with $\La \leq 2 m_\chi$.
\\

\noindent \textit{Indirect detection}

\noindent The annihilation of DM particles in regions of high density may produce visible signals in various forms. These may be $\gamma$-rays from dwarf spheroidal galaxies, collected by \textit{Fermi}-LAT, or high energy neutrinos from the annihilation of DM captured in the Sun, observed by IceCube. Furthermore, DM annihilations in the Early Universe inject energy into the primordial plasma and affect the reionisation history, which can be observed in the CMB by \textit{Planck} as changes in the optical depth. From the annihilation cross-section computed by \textsf{CalcHEP}, the indirect detection likelihoods are computed for $\gamma$-rays by \textsf{gamLike} v.1.0.1~\cite{Workgroup:2017lvb}, for neutrinos by \textsf{Capt'n General} v2.1~\cite{Kozar:2021iur} and \textsf{nulike} v1.0.9~\cite{Scott:2012mq}, and the constraints from the CMB by a combination of \textsf{DarkSUSY} and \textsf{DarkAges}~\cite{Stocker:2018avm} via \textsf{CosmoBit}~\cite{Renk:2020hbs}. As before, meaningful DM annihilations require $\La > 2 m_\chi$.
\\

\noindent \textit{Collider physics}

\noindent DM particles can be produced at LHC by proton-proton collisions. In this study we focus on a $36\,\rm{fb}^{-1}$ CMS and a $139\,\rm{fb}^{-1}$ ATLAS search, where a jet is produced via initial state radiation, leading to final states with a single jet and missing transverse energy. Collider simulations can be very CPU expensive, so we opted for pre-generating cross-section and efficiency tables, from which we interpolated the yields. We generate Monte Carlo events using \textsf{MadGraph\_aMC@NLO} v2.6.6 (v2.9.2)~\cite{Alwall:2011uj} for the CMS (ATLAS) analysis, which we later shower and hadronise using \textsf{Pythia} v8.1~\cite{Sjostrand:2007gs}. To ensure EFT validity in the collider searches we take two approaches. One where the $\slashed{E}_T$ spectrum has a hard cut-off at $\La$, and another where the spectrum has a smooth drop-off with varying slope, $(\slashed{E}_T/\La)^{-a}$. The value of $a$ depends on the specific UV completion, but since we are performing an EFT analysis, we vary $a$ as a free parameter of the model.

\section{Results}

We present the results of this study in two parts. First we show the results with a ``capped'' LHC likelihood, i.e. where the model fits the data equally or worse than the background hypothesis. And second, we allow the full LHC likelihood, where we will use both of the EFT approaches described earlier, with hard and smooth cut-offs on the $\slashed{E}_T$ spectrum.

\begin{figure*}[h]
	\centering
    \includegraphics[width=0.32\textwidth]{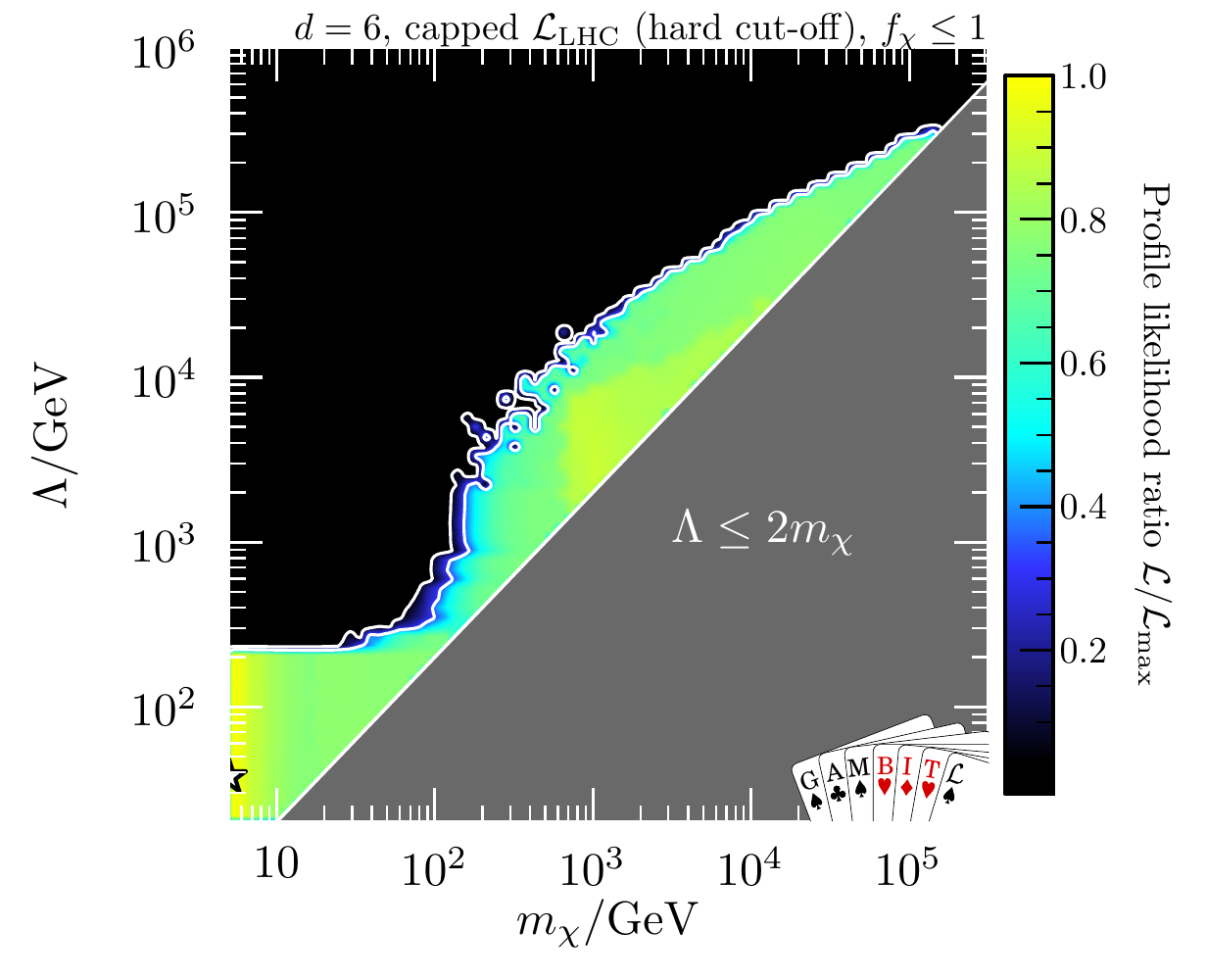}
	\includegraphics[width=0.32\textwidth]{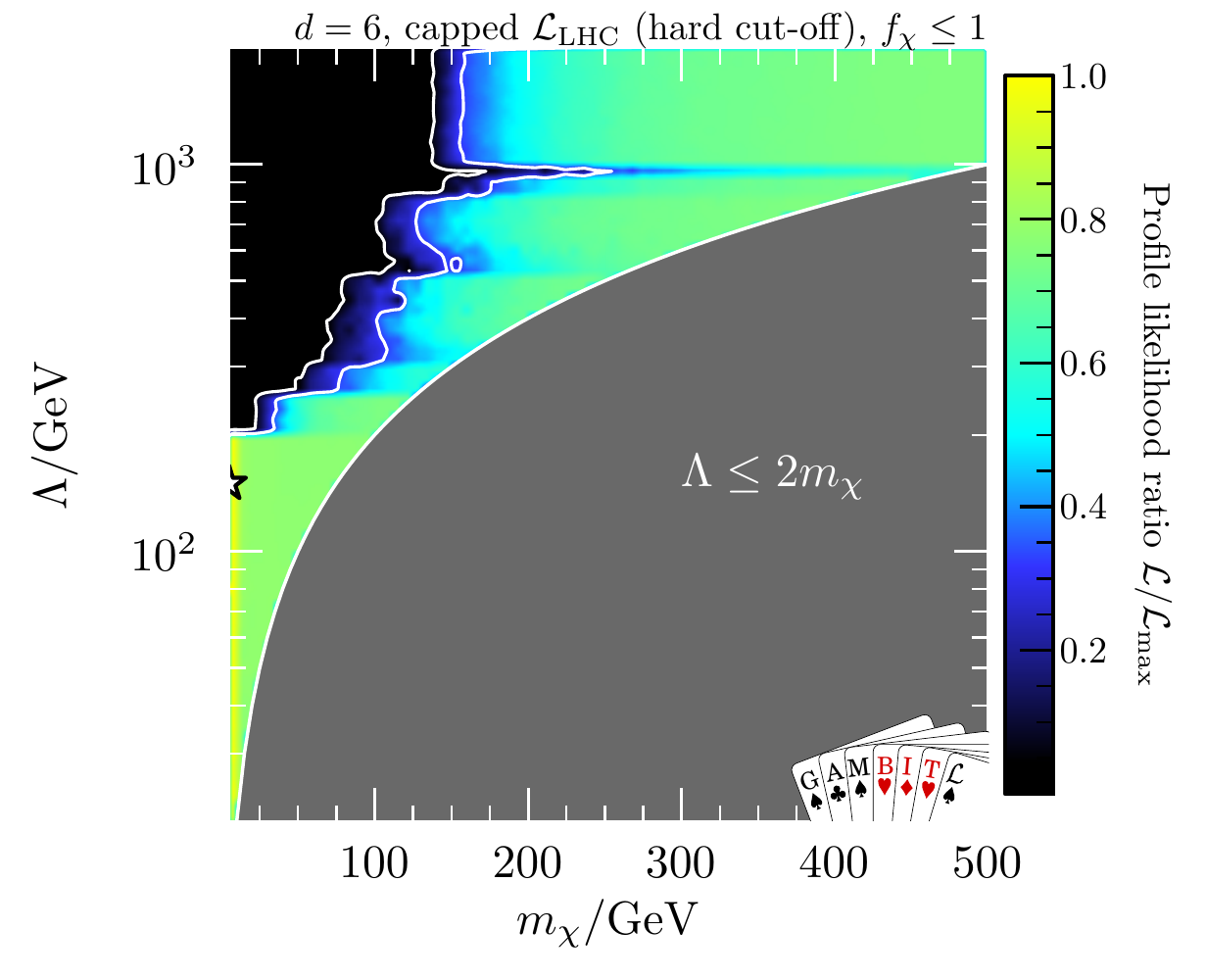}
    \includegraphics[width=0.32\textwidth]{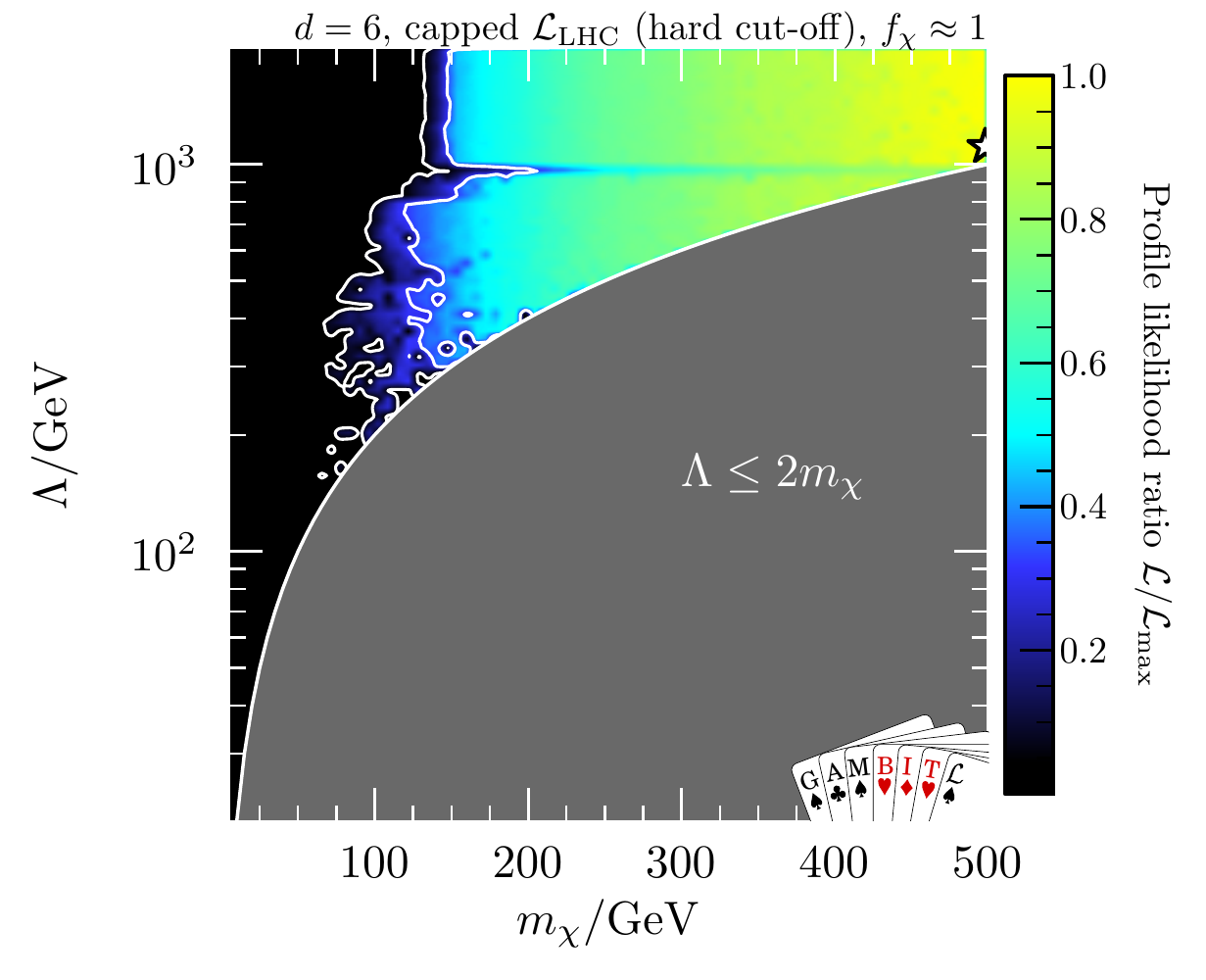}
	\caption{Profile likelihood in the full range (left) and most interesting region (centre) of the $m_\chi$--$\Lambda$ plane with dimension-6 operators, a capped LHC likelihood and the relic density as an upper bound. The right-hand plot shows the low mass region with saturated relic density. The best-fit point is indicated by the white star, the white lines indicate the 68\% and 95\% confidence level regions, and the grey region is excluded by EFT validity.}
	\label{fig:dim_6_capped_main}
\end{figure*}

Figure \ref{fig:dim_6_capped_main} (left and centre) shows the profile likelihood in the $m_\chi$ vs $\La$ plane for the $d=6$ operators. The most interesting features of these figures are the large parameter region excluded by EFT validity ($\La > 2m_\chi$), the upper limit on $\La$ at high masses due to the RD requirement, and the strongly constrained region for $\La > 200$ GeV and $m_\chi < 200$ GeV where the LHC searches are the dominant constraint. The best fit corresponds to a slight excess on the \textit{Fermi}-LAT data.

It is worth exploring the scenario where the relic abundance is exactly saturated ($f_\chi \approx 1$). This can be seen in the right panel of Figure \ref{fig:dim_6_capped_main}. It is now not possible to saturate the RD bound for low mases $m_\chi \lesssim 100$ GeV, as it is incompatible with gamma ray and CMB constraints. Because of the shrunken parameter region, the best fit point with saturated relic density predicts up to $10$ signal events at the next generation of direct detection experiments, e.g. LZ.

\begin{figure*}[h]
  \centering
  \includegraphics[width=0.32\textwidth]{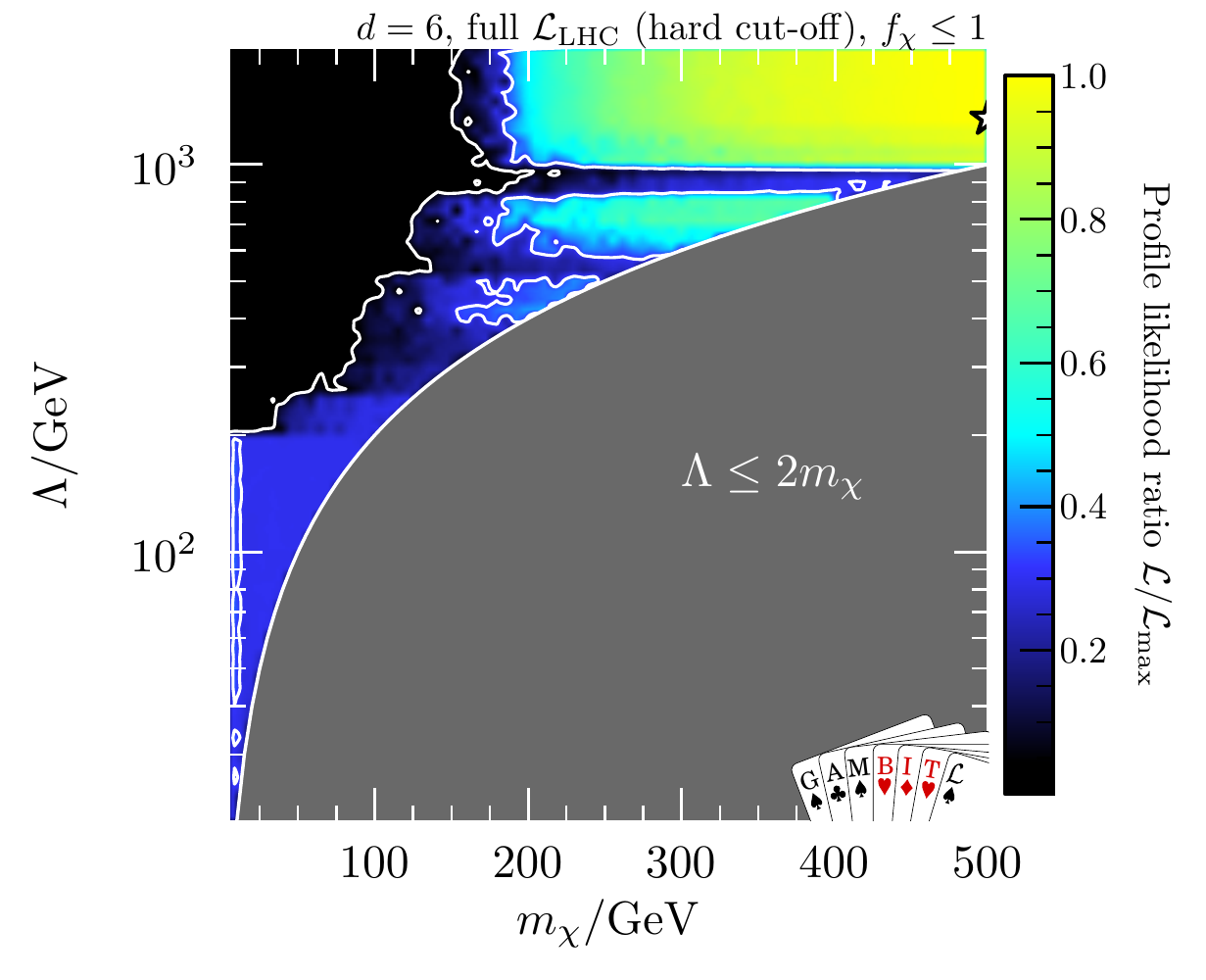} 
  \includegraphics[width=0.32\textwidth]{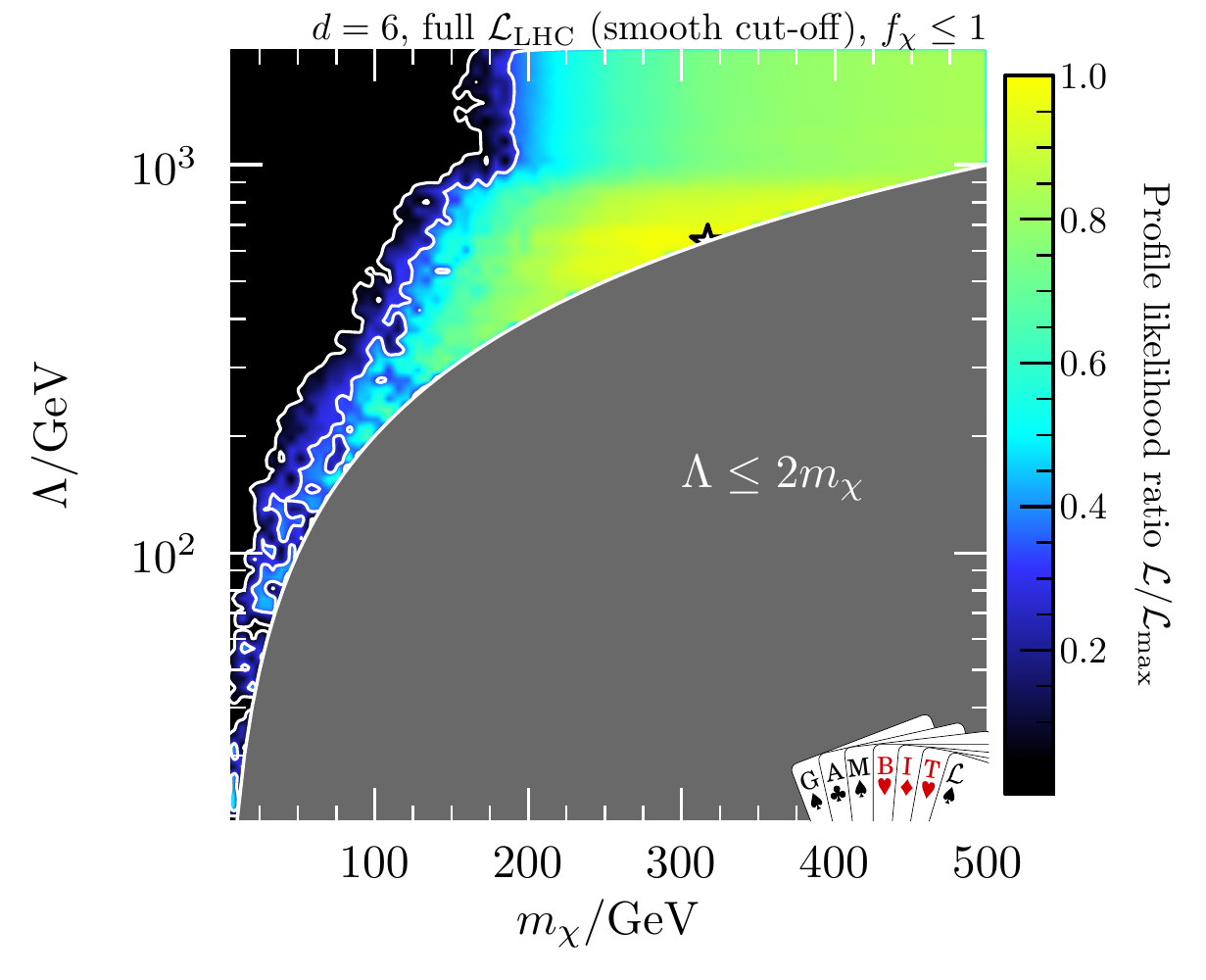}
  \includegraphics[width=0.32\textwidth]{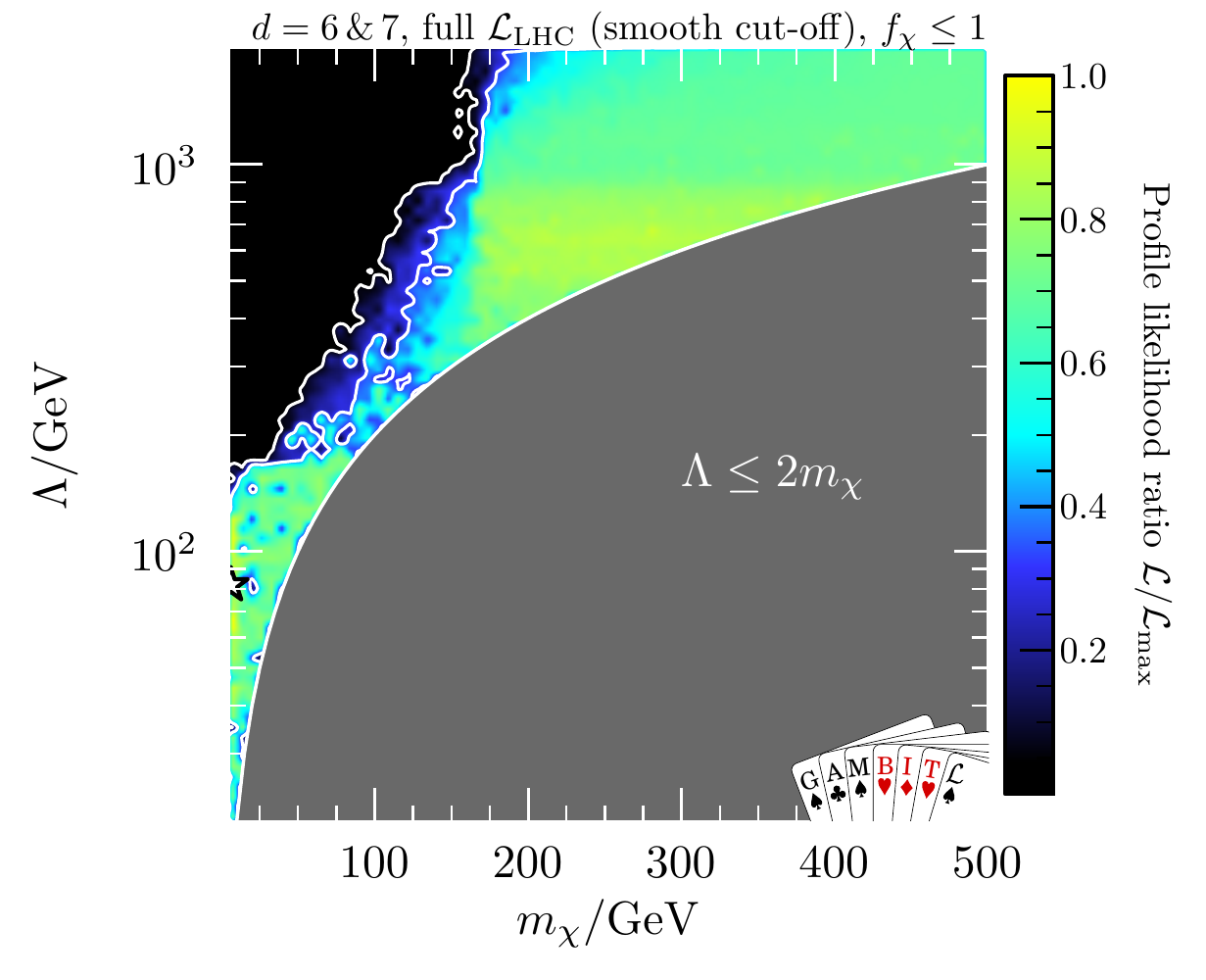}
  \caption{Profile likelihood in the $m_\chi$--$\La$ parameter plane with only $d=6$ operators (left, centre) and all $d=6$ and $d=7$ operators (right), as well as the full LHC likelihood with a hard (left) and smooth (centre, right) cut-off in the $\slashed{E}_T$ spectrum. Stars, lines are shaded regions are as in Figure~\ref{fig:dim_6_capped_main}.}
  \label{fig:dim_6_full_main}
\end{figure*}

If the LHC likelihood is included in full, it becomes the dominant constraint for $m_\chi < 500$ GeV, as can be seen in Figure~\ref{fig:dim_6_full_main}. In the $d=6$ case, there is clearly a preference now for high $\La$ values, following the bin-wise excesses in the CMS and ATLAS analyses. When using a hard cut-off on the $\slashed{E}_T$ spectrum (left), the profile likelihood shows various 1$\sigma$ regions, corresponding to values of $\La$ where the signal-to-background is maximized individually for the CMS ($\La \approx 700$ GeV) and ATLAS ($\La \gtrsim 1$ TeV) analyses. With a smooth cut-off (centre), the best fit is a combination of all excesses and thus the fit is slightly better. Although a promising feature, this slight preference over the background may be an artefact of the EFT construction, and realistic UV completions may not be able to fit individual bin excesses as the EFT approach does.

Lastly, one can study the effect of adding the dimension-7 operators. Overall the addition of the $d=7$ operators does not noticeably increase the allowed parameter space. The most notable difference is that for low $m_\chi$ it is possible to saturate the RD and avoid LHC constraints, which was impossible with only $d=6$ operators. Furthermore, with the full LHC likelihoods it is now possible to simultaneously fit the LHC excesses at high masses and the \textit{Fermi}-LAT excess at low masses, due to the increased parameter volume, as seen in the right panel of Figure ~\ref{fig:dim_6_full_main}.

\section{Conclusions}

I have presented in this conference article a summary of the first global analysis of a DM effective field theory with the full set of operators up to dimension 7. We find that there are large regions of the parameter space allowed where the scattering and annihilation cross-sections are suppresed, so that it is possible to evade direct and indirect detection constraints, while producing the right amount of relic density. We have found that constraints from LHC searches are strong for low DM masses and high new-physics scales, and that it is possible to find a slight preference for a DM signal. Nevertheless, the log likelihood ratio between the various best-fit points and the background-only hypothesis is always small, so we do not find a significant preference for a DM signal in any of our scans. This work opens the door for many subsequent studies, e.g. with lepton operators, specific UV completions or non-trivial flavour structures, among others.

\end{document}